\documentclass[%
 reprint,
 amsmath,amssymb,
 aps,
floatfix,
superscriptaddress
]{revtex4-1}

\usepackage{appendix}
\usepackage{graphicx}% Include figure files
\usepackage{dcolumn}% Align table columns on decimal point
\usepackage{bm}% bold math

\usepackage{amsmath}
\usepackage{verbatim}

\usepackage{siunitx}
\usepackage[colorlinks=true, allcolors=blue]{hyperref}
\usepackage[normalem]{ulem}
\newcommand{\diff}{\mathrm{d}}

\renewcommand{\Re}{\mathfrak{Re}}

\begin{document}

%\linenumbers %line numbers

\title{Transverse Electron Beam Shaping with Light}

\author{Marius Constantin Chirita Mihaila}
  \affiliation{University of Vienna, Faculty of Physics, VCQ, 
A-1090 Vienna, Austria}
\affiliation{University of Vienna, Vienna Doctoral School in Physics, 
A-1090 Vienna, Austria}
  \affiliation{University of Vienna, Max Perutz Laboratories, 
Department of Structural and Computational Biology, A-1030 Vienna, 
Austria}
\author{Philipp Weber} 
  \affiliation{University of Vienna, Faculty of Physics, VCQ, 
A-1090 Vienna, Austria}
  \affiliation{University of Vienna, Max Perutz Laboratories, 
Department of Structural and Computational Biology, A-1030 Vienna, 
Austria}
\author{Matthias Schneller} 
  \affiliation{University of Vienna, Faculty of Physics, VCQ, 
A-1090 Vienna, Austria}
  \affiliation{University of Vienna, Max Perutz Laboratories, 
Department of Structural and Computational Biology, A-1030 Vienna, 
Austria}
\author{Lucas Grandits}
  \affiliation{University of Vienna, Faculty of Physics, VCQ, 
A-1090 Vienna, Austria}
  \affiliation{University of Vienna, Max Perutz Laboratories, 
Department of Structural and Computational Biology, A-1030 Vienna, 
Austria}
\author{Stefan Nimmrichter}
\affiliation{Naturwissenschaftlich-Technische Fakult\"at, Universit\"at Siegen, Walter-Flex-Stra\ss e 3, 57068 Siegen, Germany}
\author{Thomas Juffmann}
\thanks{Corresponding author: thomas.juffmann@univie.ac.at}
  \affiliation{University of Vienna, Faculty of Physics, VCQ, 
A-1090 Vienna, Austria}
  \affiliation{University of Vienna, Max Perutz Laboratories, 
Department of Structural and Computational Biology, A-1030 Vienna, 
Austria}
\date{\today}

\begin{abstract}
Interfacing electrons and light enables ultrafast electron microscopy, quantum control of electrons, as well as new optical elements for high sensitivity imaging. 
Here we demonstrate for the first time programmable transverse electron beam shaping in free space based on ponderomotive potentials from short intense laser pulses. We can realize both convex and concave electron lenses with a focal length of a few millimeters, comparable to those in state-of-the-art electron microscopes. We further show that we can realize almost arbitrary deflection patterns by shaping the ponderomotive potentials using a spatial light modulator. Our modulator is lossless, programmable, has unity fill factor, and could pave the way to electron wavefront shaping with hundreds of individually addressable pixels. 
\end{abstract}

\maketitle

\section{\label{sec:level1}Introduction}

%\linenumbers

The precise control of electrons facilitated microscopes that revolutionized materials science~\cite{Courtland2018} and structural biology~\cite{Nogales2016}. One of the key technologies for this was aberration correction~\cite{HAIDER199853,rose2009historical}. While Scherzer realized~\cite{Scherzer1936} that spherical aberrations are unavoidable with static, circularly symmetric lenses, modern aberration correctors avoid this symmetry and enable both atomic-resolution imaging~\cite{HAIDER199853,rose2009historical,resolution1,resolution3} and a finite degree of wavefront engineering.
In light optics, spatial light modulators and mirror arrays allow for programmable arbitrary wavefront shaping, which has led to great advances~\cite{Rubinsztein_Dunlop_2016}, e.g.~in astronomy, deep-tissue imaging, or optical information processing. This level of control is not yet available in electron optics, even though there are applications that would demand for it, such as random-probe ptychography~\cite{Pelz2017}, efficient structure identification~\cite{Okamoto2006}, or optimized phase contrast microscopy~\cite{Juffmann2018, Bouchet2021, Koppell:21}. 

Arbitrary, but not programmable, shaping of electron waves has been demonstrated based on the transmission of electron beams through suitably sculpted material thin films~\cite{doi:10.1126/science.1198804, Uchida2010, mcmorran2011electron, SHILOH201426, doi:10.1021/acsphotonics.1c00951}. Despite inherent challenges regarding inelastic scattering, these novel electron-optical elements found applications in, e.g., aberration correction~\cite{shiloh2018spherical}, the manipulation of orbital angular momentum~\cite{doi:10.1126/science.1198804, grillo2014b}, the generation of non-diffracting Bessel~\cite{grillo2014} and self-healing Airy beams~\cite{Voloch-Bloch2013}, diffractive probe engineering~\cite{R_Harvey_2014, Ophus2016}, and contrast enhancement in biological imaging~\cite{khoshouei2017cryo,tavabi2018tunable}. Nanofabrication gave rise to beam shaping devices based on tunable electro- and magneto-static potentials. 
While some of them are tailored to specific applications, such as symmetry mapping~\cite{Guzzinati2017} or orbital angular momentum sorting~\cite{Tavabi2021}, 
others realize on-axis arrays of individually addressable electron lenses for arbitrary wavefront shaping in reflection~\cite{CARROLL20151} or transmission~\cite{VERBEECK201858, ibanez}. 
Limiting factors are the achievable transmission, unwanted diffraction from sub-apertures, charging and beam-induced deterioration.

Electron wave-packets can also be shaped by optical fields in the vicinity of matter~\cite{barwick2009photon,vanacore2018attosecond,vanacore2019ultrafast,Abajo2020,Feist2020,talebi2020strong, wang2020coherent, reinhardt2020theory, madan2022ultrafast, shiloh2022quantum} enabling attosecond coherent control of the electron wave function as well as multidimensional nanoscale imaging and spectroscopy. Although these methods do not rely on nanofabrication techniques, they still suffer from inelastic scattering and material deterioration. 

The control of electrons in free space through ponderomotive interaction with light avoids these problems. Kapitza and Dirac predicted that electrons could be diffracted at a standing light wave via a second-order process~\cite{Kapitza1933}. Early demonstrations include electron deflection~\cite{Bucksbaum1988} and diffraction~\cite{freimund,freimund2002bragg}. 
Nowadays, ponderomotive electron-light interactions are used to characterize~\cite{Siwick:05, Hebeisen:06, Hebeisen:08} and create~\cite{Kozak2018} ultrashort electron pulses, to modulate the spectrum~\cite{kozak2018inelastic} and the phase~\cite{schwartz2019laser, PhysRevLett.124.174801} of electron beams.
The challenges lie in interfacing electron optics with the required high-intensity light optics, which offers unity transmission, tunability, and does not cause inelastic scattering. 

Here, we demonstrate controllable transverse shaping of free electrons into an arbitrary density profile using the ponderomotive potential of a spatially modulated light pulse. We realize the scheme in a modified scanning electron microscope (SEM), similar to the recently proposed optical free-space modulator (OFEM) technique~\cite{PhysRevLett.126.123901}. 
Our OFEM has a pixel size of $\SI{4.3}{\mu m}$ (FWHM of focus spot), which in principle facilitates $2\pi$ phase modulation on an interaction plane of up to 1607 pixels, given the $\SI{17.2}{\mu J}$ maximum laser pulse energy in our setup. Contrary to static, cylindrically symmetric, and energy preserving electron optics, we achieve both positive and negative lensing, as has recently been proposed in a similar setup~\cite{PhysRevApplied.16.L011002}. We further can create almost arbitrary electron deflection patterns using shaped light fields.

\section{\label{sec:level2}Setup}
\begin{figure}[t]
   \centering      
   \includegraphics[width=8cm, height=6.5cm]{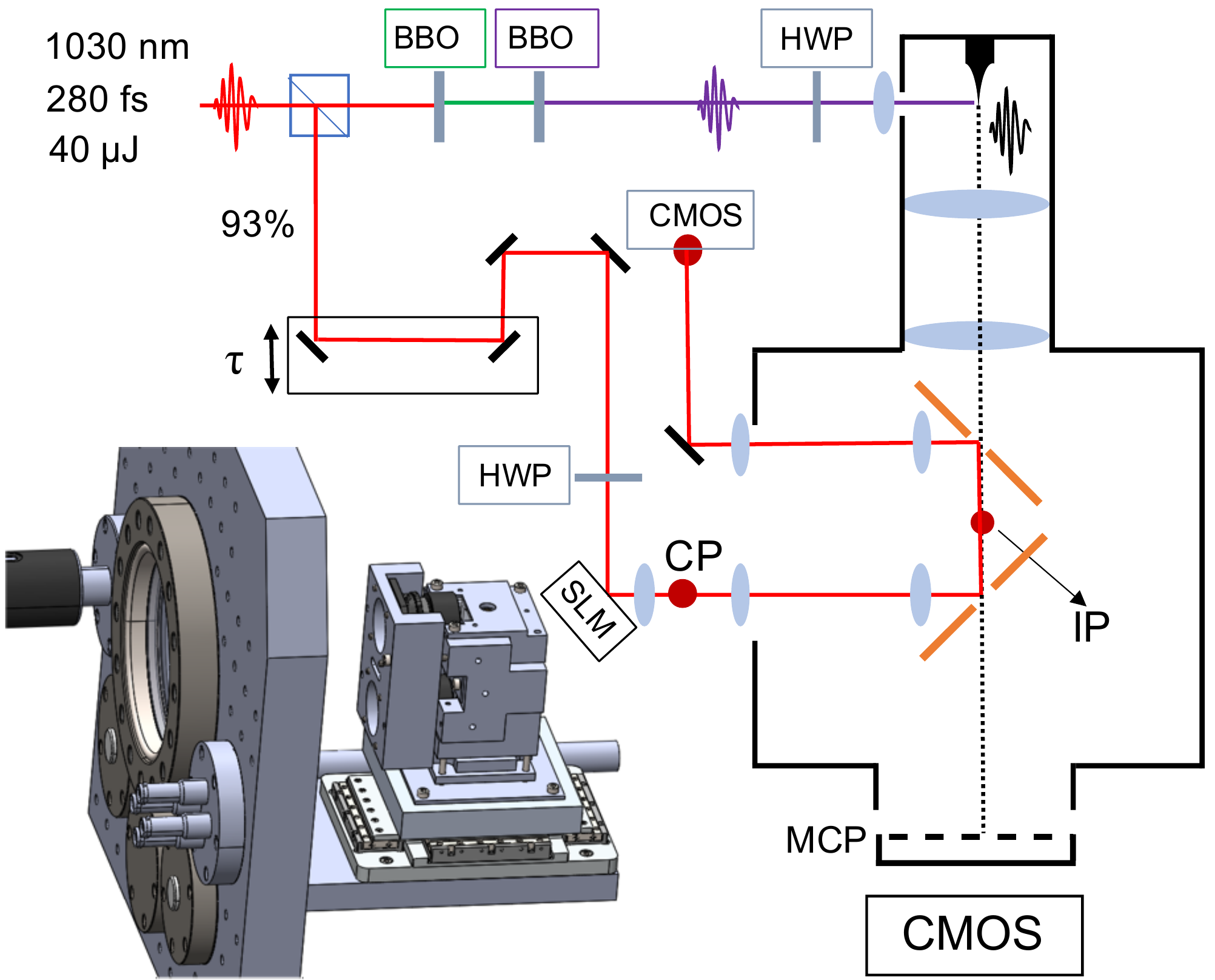} \caption{Experimental setup: Programmable pondemorotive deflection of electrons in free space. Outside of the vacuum chamber, the SLM shaped laser distribution is formed in the conjugate plane (CP). In the specimen chamber the electron beam interacts with a counter-propagating laser pulse in the interaction plane (IP). The induced electron phase modulations are proportional to the local laser pulse energy. (left) The modified SEM chamber door which allows for in- and outcoupling the shaped light fields (see also Appendix \ref{sec:B}).}
   \label{fig:setup}
\end{figure}

Experiments are carried out in a modified ultrafast scanning electron microscope (U-SEM, based on a commercial FEI XL30), as sketched in Fig.~\ref{fig:setup}. The Vienna U-SEM features a laser triggered electron source, an optical setup for shaping ponderomotive potentials, and a hole in the bottom of the specimen chamber for experiments in transmission mode. An ultrafast laser (Coherent Monaco, wavelength $\lambda = \SI{1035}{nm}$, pulse length $\Delta t = \SI{280}{fs}$, repetition rate $\SI{1}{MHz}$, pulse energy $<\SI{40}{\mu J}$) is split using a 93:7 beam splitter.

The less intense beam is used to generate the forth harmonic (UV, $\SI{258}{nm}$, $<\SI{2}{nJ}$) with help of two Beta Barium Borate (BBO) nonlinear crystals. We then focus this UV beam onto the Schottky electron gun through a 15\,cm focal length lens~\cite{PhysRevLett.96.077401}, aligning the polarization parallel to the tungsten nanotip with a half-wave plate (HWP). This creates a pulsed electron beam propagating through the column of the U-SEM. 

The more intense infrared beam creates the ponderomotive potentials (IR, $\SI{1035}{nm}$, $<\SI{17.2}{\mu J}$), after passing a delay stage that adjusts the timing ($\tau$) between IR and UV. The IR beam is then expanded to a ($1/e^2$) diameter of 7.5 mm and reflected off a water-cooled spatial light modulator (SLM, Meadowlark 1920x1152). The SLM shapes the wave-front of the laser such that the desired intensity distribution is created in the interaction plane of electrons and light, as well as in a conjugate plane used for beam characterization. We calculate the required patterns either via simple back-propagation, or via a modified Gerchberg-Saxton algorithm, which allows us to generate almost arbitrary light intensity distributions~\cite{gerchberg1972practical, Pang:16} (GSA, see Appendix~\ref{sec:A}). After passing the interaction plane, the laser is out-coupled from the U-SEM specimen chamber, which minimizes heating effects in the chamber and enables additional temporal and spatial coarse alignment (see Appendix~\ref{sec:B}). 

The in-vacuum parts of the optical system are mounted to a custom U-SEM door (see inset in Fig.~\ref{fig:setup}) that features a non-magnetic 2D piezo stage to align light and electron optics. The beams are overlapped using unprotected gold mirrors on copper substrate, which reflects the incoming light fields, but transmits the electrons through a central $\SI{1.5}{mm}$ hole. The beam shaping optics are mounted directly to the U-SEM door (see Appendix \ref{sec:B}). 
After interaction with the shaped laser pulse, the electron pulse propagates through free-space for $d=\SI{550}{mm}$. It is then detected using a micro channel plate (MCP, see Appendix~\ref{sec:C}).

\section{\label{sec:level3}Ponderomotive Phase Modulation}

Consider the paraxial wave function of an electron travelling at a relativistic velocity $v$ along the axial $z$-direction, $\psi(\bm r, t) = e^{i \gamma m_e (v z-c^2 t)/\hbar } \psi_{\perp}(x,y,t)$, with $m_e$ the electron mass, $\gamma = 1/\sqrt{1-v^2/c^2}$ the relativistic factor and $\psi_{\perp}$ the transverse electron wave shape. 
The electron shall be illuminated by a short laser pulse of moderate intensity, described by a vector potential $\bm A (\bm r, t)$ with a temporal profile extending over $10^2$ optical periods. Following~\cite{PhysRevLett.126.123901} and neglecting the transverse electron motion, the light-induced velocity modulation, as well as electron-positron mixing, we can approximate the interaction by an effective scattering phase 
$\varphi(\bm r) = -(1/\hbar) \int_{-\infty}^\infty \diff t \, U (\bm r + vt \bm e_z,t)$
imprinted onto the wave function, with the ponderomotive potential
\begin{equation}
U (\bm r,t) = \frac{e^2}{2m_e \gamma } \left[ A_x^2(\bm r, t) + A_y^2(\bm r, t) + \frac{A_z^2(\bm r, t)}{\gamma^2} \right].
\end{equation}

In our setup, the light intensity patterns are formed with optics of moderate numerical aperture ($\text{NA}<0.2$), which implies $A^2_z(\bm r, t) \approx 0$. Hence, the potential does not depend on the light polarization. The small NA also entails that the depth of focus is larger than both the distance the electron propagates during the interaction and the dispersed axial width of the electron wave packet~\cite{Kozak2018b}. This is in contrast to the derivation in~\cite{PhysRevLett.126.123901}, where longer laser pulses were assumed, such that the electron would also interact with out-of-focus light. In our setting, each electron interacts with the total laser pulse in focus, making the scheme more energy efficient. This allows us to separate the temporal laser pulse profile, specified by a function $u(t)$, from the transverse spatial profile, specified by $g(x,y)$ and polarized along, say, the $x$-axis. For a laser pulse that co-propagates (upper sign) or counter-propagates (lower sign) with the electrons, the real-valued amplitude reads as
\begin{equation}
    \bm A(\bm r,t) \approx - \frac{\mathcal{E}_0}{\omega_L} \bm e_x g(x,y) u \left( t \mp \frac{z}{c} \right) \sin \left(\omega_L \left( t\mp \frac{z}{c}\right)\right),
\end{equation}
with $\omega_L = 2\pi c/\lambda_L$ the central laser pulse frequency. For typical pulses much longer than the optical period, the field strength $\mathcal{E}_0$ can be expressed in terms of the total pulse energy $E_L$ flowing through the $xy$-plane as
\begin{equation}
    E_L \approx \frac{c\varepsilon_0}{2} \mathcal{E}_0^2 \int_{-\infty}^\infty \diff t \, u^2(t) \int \diff x\diff y \, g^2 (x,y).
\end{equation}
Putting everything together, we arrive at a phase modulation of the transverse wavefunction,
\begin{equation}
    \varphi(x,y) \approx -\frac{\alpha}{2\pi (1\mp \beta)} \frac{E_L}{E_e} \frac{\lambda_L^2 g^2(x,y)}{\int\diff x\diff y\, g^2(x,y)}, \label{eqn:phase_shift}
\end{equation}
with $\alpha$ the fine structure constant, $\beta=v/c$, and $E_e = \gamma m_e c^2$ the relativistic energy of the electron. We see that the phase is proportional to the ratio of pulse and electron energy times the normalized spatial light profile, while it does not depend on the temporal envelope of the laser pulse. 
See Appendix \ref{sec:D} for a detailed derivation.

\section{\label{sec:level4}Results}

\begin{figure}[htp]
  % \centering      
   \includegraphics[width=8.0cm]{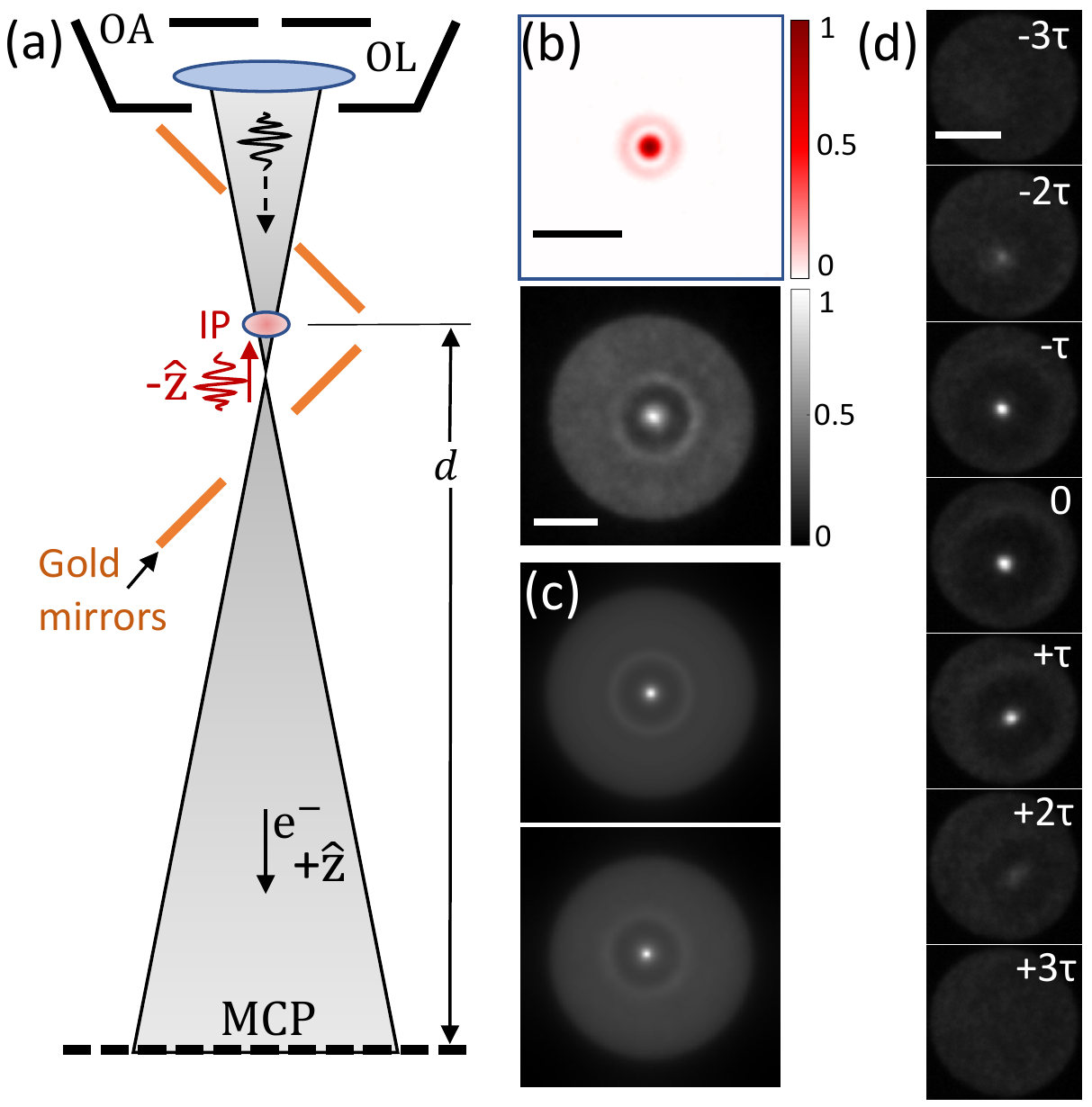}
   
 \caption{Ponderomotive electron-light interaction: 
(a) An SEM objective lens (OL) focuses the electron pulse to a cross-over slightly below the interaction plane. The objective aperture (OA) is set to $\SI{100}{\mu m}$. (b) (top) Light intensity distribution (scalebar $\SI{93}{\mu m}$) before being demagnified (5x)  to the interaction plane.(bottom) Observed electron intensity distribution (scalebar $\SI{1}{mm}$) at the detector plane. (c)~wave optics (top) and ray optics (bottom) simulations of the detected electron beam shown in (b). (d) Detected electron distributions (scalebar $\SI{1.6}{mm}$) when delaying the laser pulse with respect to the electron pulse by multiples of $\tau =\SI{666}{fs}$. The delay causes the interaction to take place in a plane where the light is not focused.}
 \label{fig:temporal}
\end{figure}

Figure \ref{fig:temporal}(a) shows the experimental configuration with a converging ($\SI{30}{keV}$) electron beam. Figure \ref{fig:temporal}(b)(top) shows the measured focused light intensity before being demagnified (5x) to the interaction plane. The focus spot of the light pulse in the interaction plane has a FWHM of~$\SI{4.3}{\mu m}$, which represents the diffraction limited minimum spot size in our setup. Figure \ref{fig:temporal}(b)(bottom) shows the electron intensity distribution observed at the MCP for a laser pulse energy of $\SI{6}{\mu J}$, an electron beam radius in the interaction plane of $\rho_0=\SI{17.5}{\mu m}$, and a distance from interaction plane to cross-over of $d_c=\SI{5.9}{mm}$.

Figure~\ref{fig:temporal}(c) compares these experimental results to wave optics (top) and ray tracing (bottom) simulations, yielding good agreement. In the wave optics model, we assume a spherical wavefront of the incoming converging beam in Fresnel approximation and compute the ponderomotive phase Eq.~\eqref{eqn:phase_shift} for a circularly symmetric fit to the measured light distribution in (b).
In the ray optics simulation, we treat the electrons as ballistic particles and hence the ponderomotive interaction as a position-dependent transverse momentum kick~\cite{PhysRev.150.1060} by $\delta \bm p_{T}=\hbar \nabla \varphi(x,y)$. Details of both methods can be found in Appendix \ref{sec:E}. They yield similar results for the high laser intensities considered here; a $\SI{6}{\mu J}$ laser pulse amounts to a phase shift in the beam center of $\SI{950}{rad}$. 
Small differences due to diffraction could not be observed due to the finite spatial resolution of the detector (see Appendix \ref{sec:C}).

Figure \ref{fig:temporal}(d) demonstrates how the electron deflection depends on the timing of the light pulse, varied in steps of $\tau=\SI{666}{fs}$ relative to the electron pulse (at $\SI{6}{\mu J}$ pulse energy). With no timing overlap (top and bottom), we detect a homogeneous electron intensity distribution. At perfect overlap (center) we see a clear signature of the ponderomotive interaction. A timing mismatch of $\pm \tau$ hardly affects the observed electron distribution, as it displaces the interaction plane by $\delta z= v\tau/(1+v/c)=\SI{51}{\mu m}$. Such displacements only lead to slight changes of the light intensity distribution (See Appendix \ref{sec:B} for details). 

Considering previous electron pulse length measurements performed in a similar system~\cite{Kozak2018b}, we can therefore assume that, at perfect timing overlap, all electrons see almost the same transverse light intensity distribution. For timing mismatches of $\pm 2\tau$, the electrons interact with a more defocused light pattern, and the ponderomotive signature is weaker. 

\begin{figure}[]
   \centering      
   \includegraphics[width=\columnwidth]{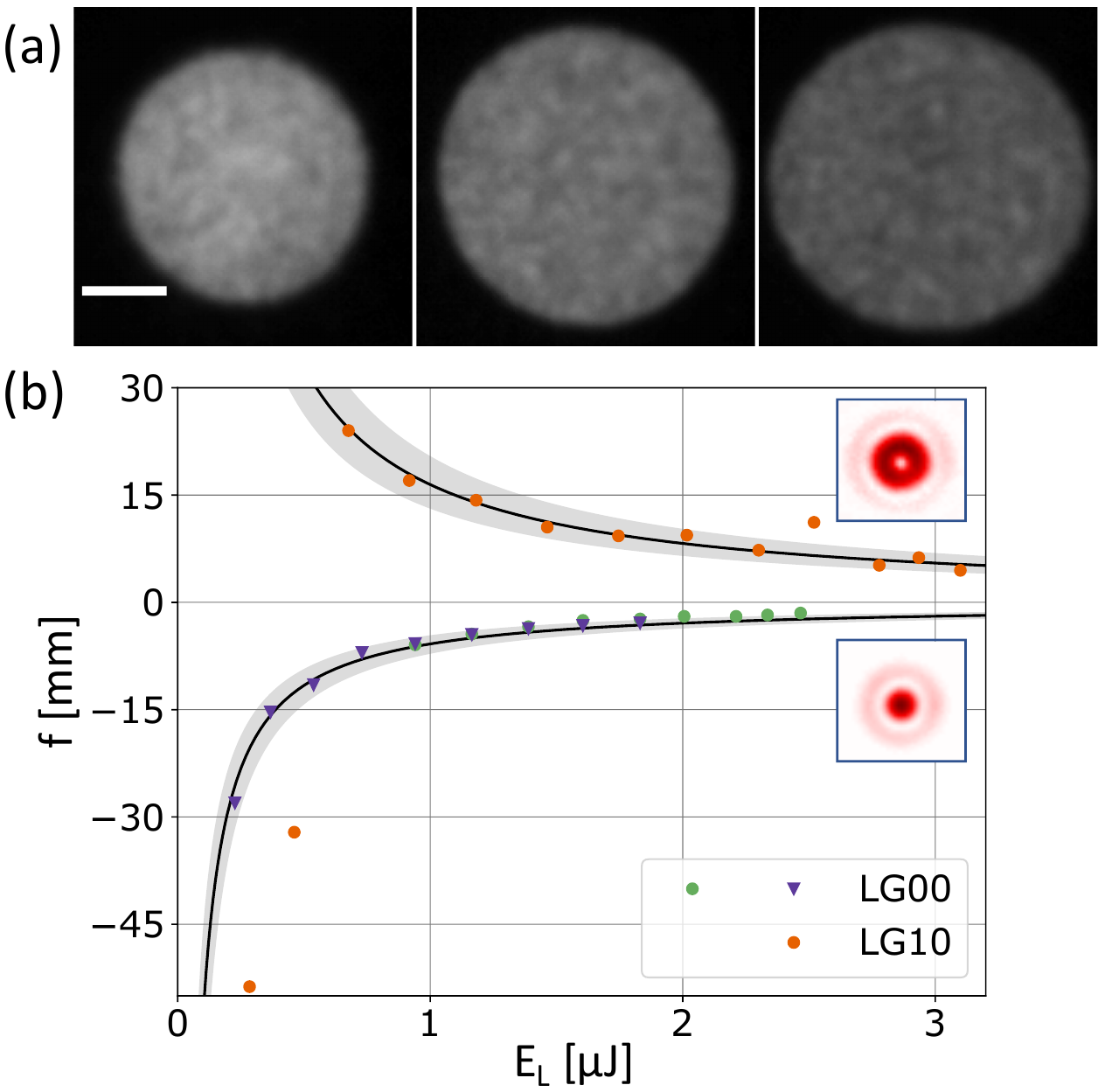}
   
 \caption{Lensing using ponderomotive potentials: 
(a)~Detected electron distribution with concave (left, LG00), no (center), and convex (right, LG10) ponderomotive lens (scalebar $\SI{1}{mm}$). (b)~Focal length as a function of laser pulse energy for LG00 (green and violet) and LG10 (orange) mode. The inset shows the respective light intensity distributions, measured in a plane conjugate to the interaction plane.}  
 \label{fig:lensing}
\end{figure}

Next, we show that ponderomotive potentials can realize both convex and concave lenses. This is in stark contrast to static, cylindrically symmetric, and electron energy preserving electron optics that can only realize a convex lens, if there are no charges on the optical axis~\cite{Scherzer1936}. A convex (concave) lens applies a phase shift to an incoming beam that decreases (increases) quadratically with distance $\rho$ from the optical axis, $\phi(\rho)=\mp \pi\rho^2/\lambda_e f$, with $\lambda_e=h/\gamma m_e v$ the deBroglie wavelength of the electron. This can be realized close to the center of focused Laguerre-Gaussian (LG) beams with radial index $p=0$. For an azimuthal index $l=0$ (LG00), the Taylor expansion of the spatial distribution is given by $g^2_{00}(\rho) \approx 1 - 2\rho^2/w_{00}^2 + \mathcal{O}(\rho^4/w_{00}^4)$, which corresponds to a concave lens. For $l=1$ (LG10) one obtains $g^2_{10} \approx \rho^2/w_{10}^2 - \mathcal{O}(\rho^4/w_{10}^4)$, which corresponds to a convex lens. In both cases, $w$ denotes the beam waist of the respective mode. 

Figure~\ref{fig:lensing} demonstrates controllable ponderomotive lensing.
The central panel in (a) shows the detected electron distribution without laser interaction, the left (right) panel shows the detected intensity upon interaction with an LG00 (LG10) mode at a pulse energy of $\SI{2.5}{\mu J}$ ($\SI{3.1}{\mu J}$).
The experimental LG light intensity distributions are shown as insets in Fig.~\ref{fig:lensing}(b), as measured in the conjugate plane at a magnification of $5\times$.
The converging electron beam illuminates the central part of these distributions, where the quadratic approximations are valid ($R\sim\SI{0.8}{\mu m}$, $d_c\sim\SI{241}{\mu m}$).
We deduce the focal length of the ponderomotive lens from the radius $\rho_{s}$ of the measured electron distribution using a ray optics model, $f = d\rho_0 / (\rho_0-\rho_{s}+d\theta_{\rho,0})$, 
with $\rho_0$ and $\theta_{\rho,0}$ the maximum radius and convergence angle of the electron beam in the interaction plane and $d$ the distance from the interaction plane to the screen (see Appendix \ref{sec:E} and \ref{sec:F}). The focal length is given by 

\begin{align}
    f(E_L) &= \pm (1\mp \beta) \frac{\pi^3 w^4}{2 \alpha \lambda_L^2 \lambda_e} \frac{E_e}{E_L}
    \label{eqn:focal_length_main}
\end{align}
Figure~\ref{fig:lensing}(b) shows this deduced focal length as a function of laser pulse energy for the convex (orange) and concave (blue and green for two independent data runs) lens. The solid line is calculated according to Eq.~\eqref{eqn:focal_length_main}, where we get $E_L$ and $w$ from fitting the measured light distributions (see Appendix \ref{sec:F}). The gray-shaded areas denote a $5\%$ uncertainty of the width that we estimate from the fit. We see good agreement between theory and experiment, except for two data points taken at low pulse energies. When the observed electron beam width is close to $\rho_{s}(E_L)\approx \rho_0+d\theta_{\rho,0}$, alignment is difficult, and small errors can lead to a sign change in $f(E_L)$. At higher pulse energies, we see that we can reach focus lengths down to $\SI{1.6}{mm}$, comparable to those found in modern TEMs.
 
\begin{figure}[t]
   \centering      
   \includegraphics[width=\columnwidth]{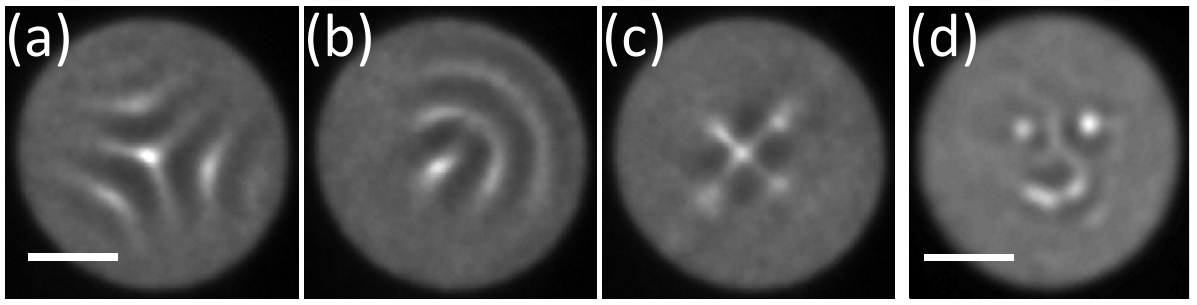}
   
 \caption{Programmable electron beam shaping: detected electron distribution after interaction with shaped light fields. For (a) to (c) the SLM was used to add vertical trefoil, coma, or astigmatism (scalebar $\SI{1.1}{mm}$) to the incoming light field. In (d) a Gerchberg-Saxton-algorithm was used to shape the light field into a distribution resembling a Smiley, which also appears in the detected electron distribution (scalebar $\SI{1.2}{mm}$).
}
 \label{fig:SLM}
\end{figure}

In Fig.~\ref{fig:SLM}, we demonstrate programmable electron beam shaping at a laser pulse energy of $\SI{6}{\mu J}$. First, we use the SLM to add trefoil, coma, or astigmatism to the optical beam leading to the characteristic electron deflection patterns in panels (a) to (c), respectively. The electron beam settings are the same as in Fig.~\ref{fig:temporal}(b)(bottom). Finally, Fig.~\ref{fig:SLM}(d) shows a smiley imprinted onto the electron beam ($d_c=\SI{8.4}{mm}$, $\rho_0=\SI{24}{\mu m}$). While for all other figures, we calculate the phase pattern applied to the SLM using back-propagation from the conjugate plane to the SLM, we here use a GSA with 20 iterations to converge to the desired pattern (see Appendix \ref{sec:A}).

\section{\label{sec:level5}Discussion}
 Our experiments show that we can deflect electron beams almost arbitrarily by shaping ponderomotive potentials. The spatial shapes are only limited by diffraction as well as by artefacts induced by the hole in the $\SI{45}{degree}$ mirror;
 our currently used diameter of $\SI{1.5}{mm}$ facilitates alignment, but could be reduced in future setups.
 Smaller features would be possible using light of shorter wavelength or higher numeric aperture. However, such measures also reduce the depth of field of the shaped intensity distributions, which sets constraints on the pulse lengths of electrons and light. 

Although we operated our setup in the high-power limit, typical wave-front shaping applications demand lower laser pulse energies.
At $\SI{30}{keV}$, the energy density required to achieve a ponderomotive phase shift of $2\pi$ is $\SI{0.58}{\nano J}/\SI{}{\mu m^2}$. Our system, which offers an IR pulse energy in the interaction plane of up to $\SI{17.2}{\mu J}$, could therefore support the shaping of an area with a radius $R\lesssim \SI{97}{\mu m}$, corresponding to $N\sim(2*R/\text{FWHM})^2=1607$ individual pixels.
 
A miniaturized version of our setup could also be implemented in ultrafast electron microscopes~\cite{zewail20064d,berruto2018laser,da2018nanoscale}, where it could be used for aberration correction \cite{VERBEECK201858, shiloh2018spherical}. It has been shown that 100 addressable pixels could effectively correct spherical aberration~\cite{VERBEECK201858}. With similar light focusing capabilities as described here, $100$ addressable pixels with $2\pi$ phase modulation capability can be realized by shaping potentials on a disc with $\rho_0\simeq \SI{25}{\mu m}$. Modern TEMs provide the required level of coherence across such an area~\cite{FEIST201763}. In a counter-propagating mode, a total pulse energy of $\SI{2.2}{\mu J}$ would suffice for $\SI{300}{keV}$ electrons, which is readily available at repetition rates of tens of MHz.

\section{\label{sec:level6}Conclusion and Outlook}
 In this work, we have demonstrated programmable transverse beam shaping of free electron beams.
 The technique is based on ponderomotive potentials from high intensity laser pulses, which are shaped using a spatial light modulator.
 We have shown that we can create both a convex and a concave lens, the latter being impossible with traditional electron optics.
 We achieve focal lengths down to~$\SI{1.6}{mm}$, which is comparable to the focal length of an objective lens in a TEM.
 Future experiments might use consecutive laser pulses for realizing complex electron optical systems made from light, similar to schemes used for the creation of attosecond pulse trains~\cite{Kozak2018}. We have further demonstrated our ability to imprint complex patterns and arbitrary images onto an electron beam, with applications in fast and shaped beam blanking.
 Our work paves the way to lossless electron wavefront shaping with hundreds of individually addressable pixels. This could have applications in aberration correction~\cite{shiloh2018spherical, VERBEECK201858} of pulsed electron microscopes~\cite{Zewail2006}, if the optical modulator is installed in a suitable plane within the microscope. Our shaping technology avoids scattering, sub-aperture diffraction, and loss, the latter being especially important for pulsed electron microscopes, where the small electron current leads to long acquisition times, and consequently to stability issues. Further studies are required to show whether ponderomotive electron beam shaping can reach the noise levels present in current aberration correctors \cite{uhlemann2013thermal, linck2016chromatic}. Most importantly, programmable ponderomotive shaping could be used for applications intractable with traditional aberration correctors: These include the creation of exotic beams~\cite{grillo2014, Voloch-Bloch2013}, probe engineering~\cite{R_Harvey_2014, Ophus2016}, pulsed ponderomotive phase plates for phase microscopy~\cite{schwartz2019laser}, and other adaptive measurement techniques~\cite{Okamoto2006, Juffmann2018, Koppell:21} that maximize the extracted information in a sample-specific way.

\begin{acknowledgments}
We thank F. Balzarotti, P. Haslinger, and M. Arndt for fruitful discussions. We further acknowledge support from the Electron Microscopy Facility team
at Vienna BioCenter Core Facilities. This project has received funding from the European Research Council (ERC) under the European Union’s Horizon 2020 research and innovation program (Grant Agreement No 758752). The authors declare no conflicts of interest. 
\end{acknowledgments}

\appendix

\section{\label{sec:A}Creating arbitrary light intensity distributions}

\begin{figure}[hb]
   \centering      
   \includegraphics[width=8cm, height=4cm]{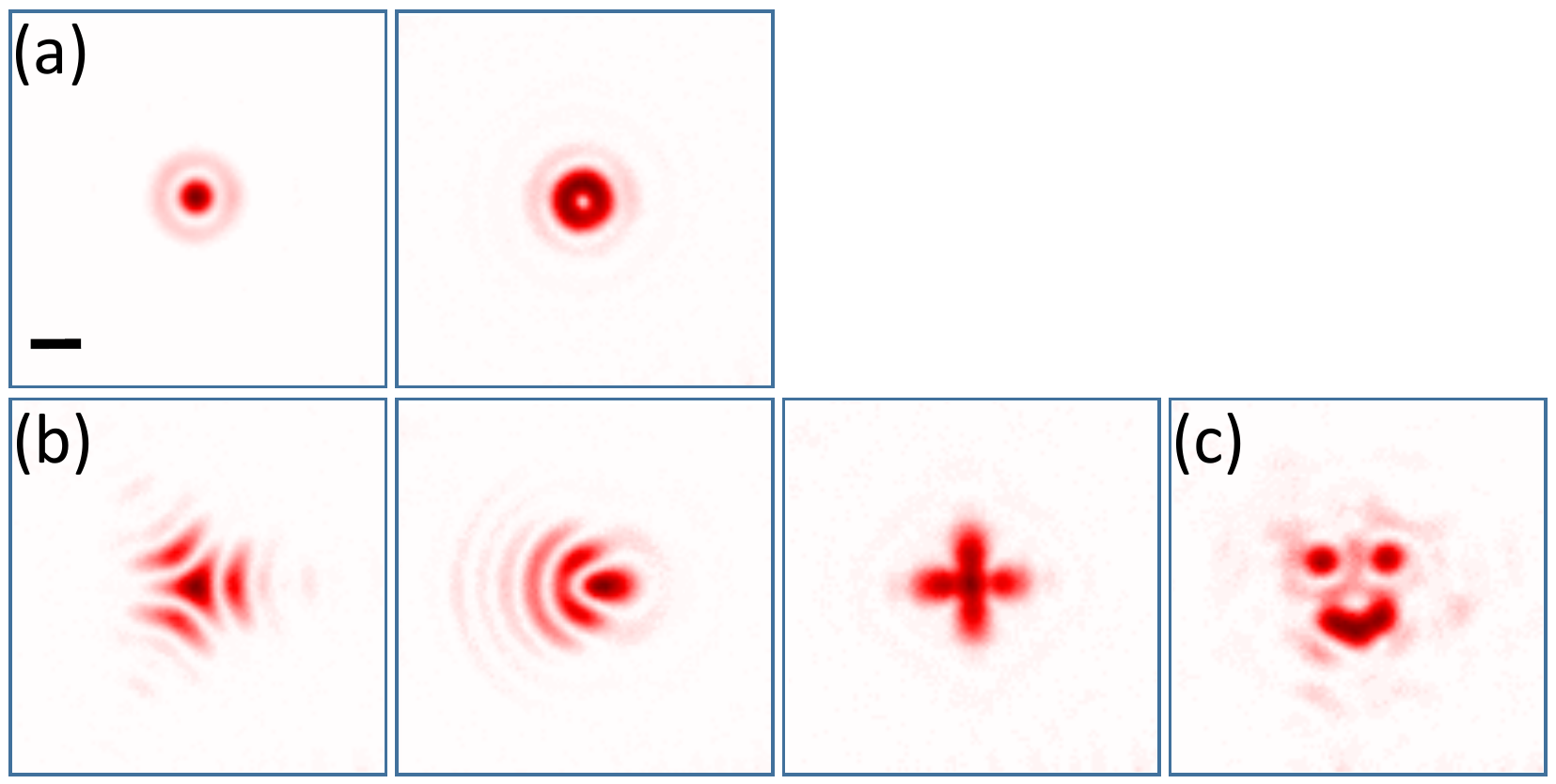}  
   \caption{Measured target intensities with the CMOS (scalebar $\SI{33}{\mu m}$) in the conjugate plane before being demagnified (5x) to the IP :(a) For the lens (b) Programmable intensity modulation (c) Arbitrary patterns, smiley.}
   \label{fig:1FFTimages}
\end{figure}
 
\begin{figure}[t]
   \centering      
   \includegraphics[width=8cm, height=6cm]{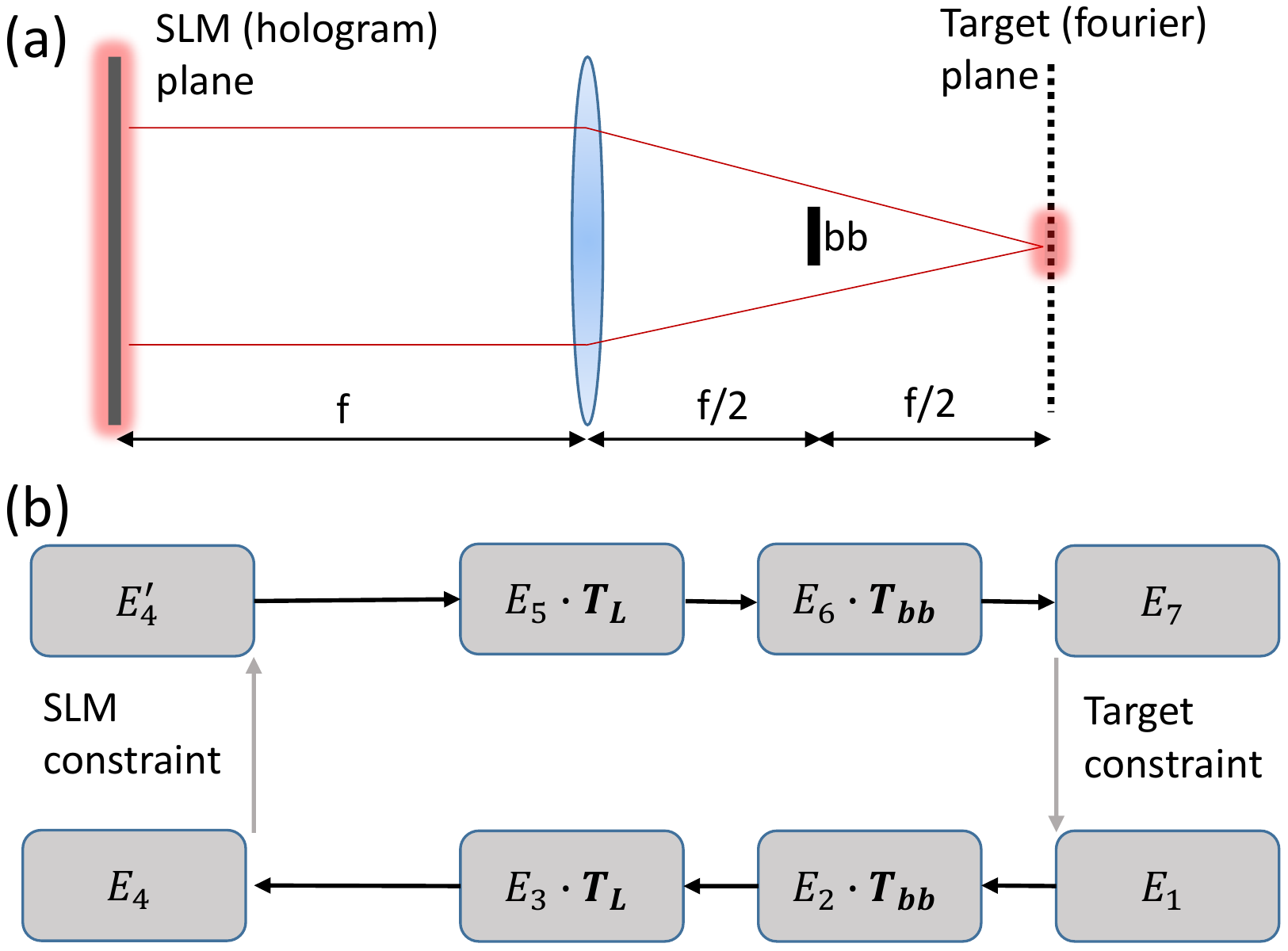}  
   \caption{(a) Optical setup used to generate the desired laser intensity distribution. In order for the electrons to pass through the optical system, a 1.5 mm hole was drilled in the mirrors before and after the interaction plane. To minimize the laser intensity losses at the hole and hence, to facilitate diffraction around it, a beam block has been implemented in the algorithm. (b) The schematic of the adapted iterative GSA. For propagating the electric field from one plane to another, the Fresnel transfer function has been used. This allows for introducing additional elements, like beam block, in the iterative algorithm.}
   \label{fig:GSA}
\end{figure}

 Figure~\ref{fig:1FFTimages} shows the light intensity patterns used to shape the electron beam in Figs.~\ref{fig:temporal}, \ref{fig:lensing}, and \ref{fig:SLM}. The images were taken outside of the SEM, in a plane conjugate to the interaction plane (target plane). The optics that image these patterns into the interaction plane also demagnify them by a factor 5x. To calculate the SLM phase masks that create these intensity distributions, two different algorithms are used. Both are based on Fresnel propagation. They take the finite aperture of our system into account, as well as the hole in the 45 degree mirror, which is simulated as a beam block (see Fig.~\ref{fig:GSA}(a)). The latter prevents loss within the hole, and therefore avoids heating in the SEM chamber. While it does affect the quality of the resulting intensity distributions, the effect is acceptable for our proof-of principle experiments. The diameter of the hole in the current implementation is $\SI{1.5}{mm}$.

The first algorithm is used for the Gaussian and the Laguerre-Gaussian focal spots in Fig.~\ref{fig:1FFTimages}(a), as well as for the intentionally aberrated focal spots in Fig.~\ref{fig:1FFTimages}(b). A Gaussian focal field is back-propagated from the target plane to the SLM. The resulting phase distribution ($\phi_G$) applied to the SLM yields the approximately Gaussian intensity distribution in Fig.~\ref{fig:1FFTimages}(a). Additional phase patterns are then added in order to create the Laguerre-Gaussian focal spot $\phi_{LG10}=
\arg (x+iy)$, as well as the spots affected by trefoil $\phi_{Z_3^{-3}}=\pi(-x^3+3xy^2)$, coma $\phi_{Z_3^{-1}}=\pi(-2x+3x^3+3xy^2)$ and astigmatism $\phi_{Z_2^2}=\pi(-x^2+y^2)$. 
Here, $x$ and $y$ are the cartesian coordinates on the SLM in units of $R_\text{SLM}=\SI{5.3}{mm}$, which denotes the radius of the area that is modulated by the SLM. $Z_n^m$ signifies the Zernike polynomials characterizing the specific aberration.  

The second algorithm is used for arbitrary patterns, like the smiley in Fig.~\ref{fig:1FFTimages}(c). It is based on the Gerchberg-Saxton algorithm (GSA)~\cite{gerchberg1972practical}, as schematically described in Fig.~\ref{fig:GSA}(b). The algorithm is initiated with the target amplitude and a constant phase ($E_1$). The field then iteratively propagates between the SLM and target plane, taking into account the transmission function of the beam block ($T_{bb}$) and the lens ($T_L$). In the SLM (target) plane, the amplitude of the calculated field is replaced by the amplitude of the incoming (desired) field. For the smiley, 20 iterations showed the best experimental results. While more iterations improved results in the simulation, the experimental results deteriorated due to the emergence of speckle~\cite{Pang:16}. 

In both algorithms the desired pattern is offset from the optical axis. This leads to an additional blazed grating on the SLM, which effectively separates the desired intensity pattern from the unmodulated beam that is reflected off the SLM~\cite{Jesacher:07}. On top of that, the positioning of the laser beam with respect to the electron beam can be adaptively optimized. Finally, the aberrations of our optical setup are partially compensated by an additional phase mask that was found by optimizing the experimentally obtained Gaussian focus spot. 

\section{\label{sec:B} Experimental Setup and Alignment Procedure}
\begin{figure}[htp]
    \centering
    \includegraphics[width=8.5cm, height=7.6cm]{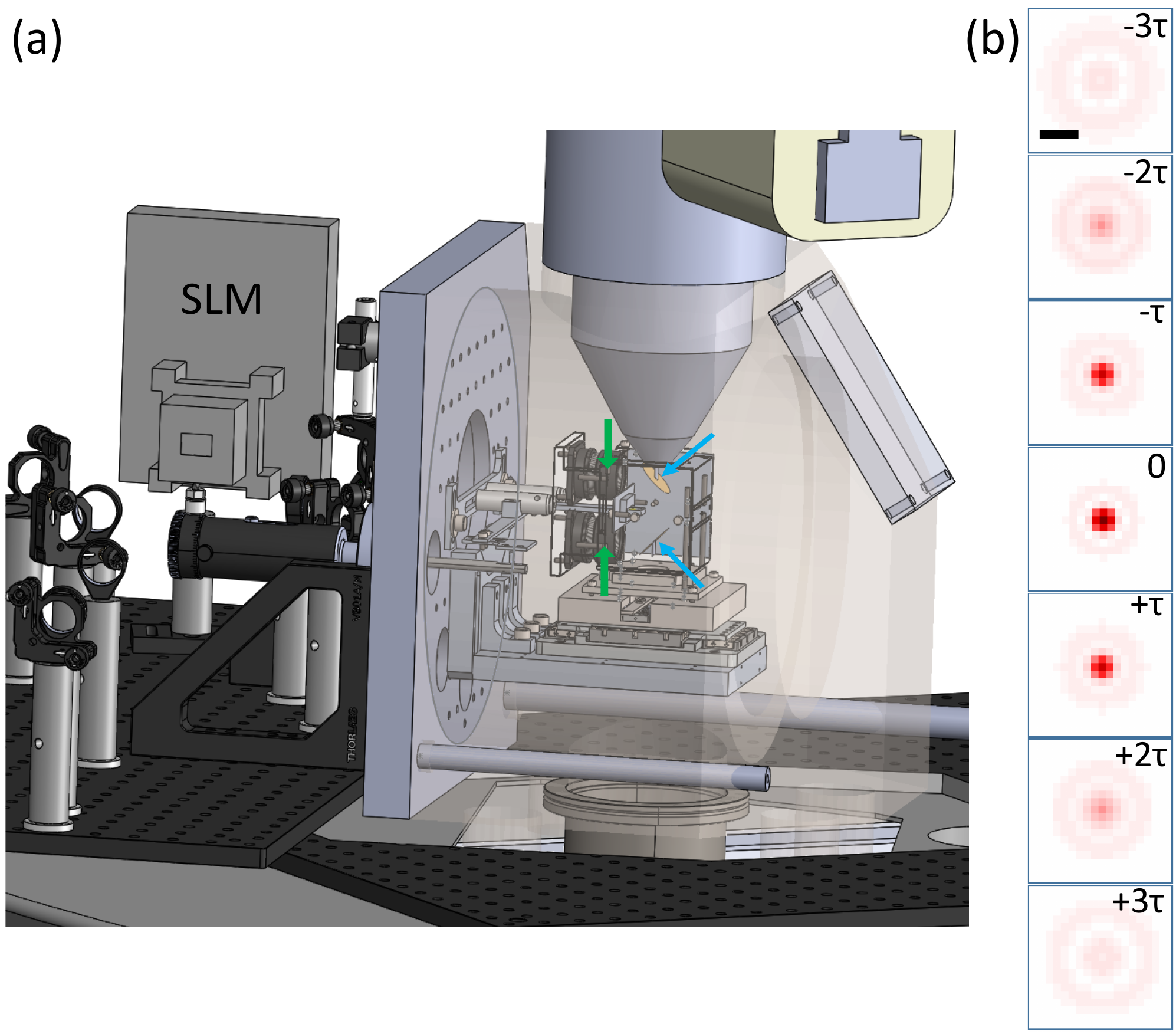}
    \caption{(a) 3D drawing of the experimental setup, side view. (b) Series of simulated ponderomotive potentials (scale bar: $\SI{45}{\mu m}$) in the conjugate plane after temporally displacing the laser pulse by multiples of $\tau$. The same linear scale was used for all images.}
    
    \label{fig:setup_side}
\end{figure}
Figure~\ref{fig:setup_side}(a) shows the custom SEM door that carries the experiment. Outside the vacuum chamber, an optical breadboard is mounted directly to the SEM door (Fig.~\ref{fig:setup_side}(a), left), and holds the SLM optics for shaping the laser pulses. An active beam pointing stabilization system (Aligna, TEM-Messtechnik) is used to counter-act pointing instabilities (25 µrad according to the specifications of our laser), as well as vibrations and drifts between the laser system and the SEM. Inside the chamber (Fig.~\ref{fig:setup_side}(a), right), the setup for overlapping light and electrons is placed right under the objective lens of the SEM. 
The whole system is attached to the SEM door, and the optics are mounted on a 2D piezo stage to facilitate alignment with the electron beam. Light enters and exits the setup through a CF 100 window in the SEM door. The green arrows mark the in-coupling and out-coupling lenses, and the blue arrows mark the mirrors that overlap the two beams.

We establish the spatial and temporal overlap between electrons and light in a multi-step process. First, a fast timing scintillator (BC-408) is placed in the interaction plane and imaged onto a CMOS camera outside of the vacuum. This is achieved using an additional manually operated mechanical stage designed to fit into the narrow space between the two lenses. This allows finding spatial overlap between the transmitted IR light and the photons triggered by the incoming electron beam. These photons are further time-stamped with a fast single photon detector (SiPM, KETEK PM1125)~\cite{6882845} to reduce the timing mismatch to within $\sim \SI{200}{ps}$. Second, overlap in the interaction plane is established doing pump-probe measurements with a copper TEM grid placed in the interaction plane~\cite{dwyer2006femtosecond}.
Finally, the copper grid is removed and the overlap between IR laser and electrons in free space is confirmed by measuring electron deflections caused by strong pondemorotive potentials. Fig.~\ref{fig:setup_side}(b) shows simulated ponderomotive potentials as they were used for the temporal and spatial alignment in Fig.~\ref{fig:temporal}(d).

\section{\label{sec:C}Electron detection}

For electron detection, we use a microchannel plate detector (Tectra GmbH, MCP-025-D-L-F-V-P43-6MI) in chevron configuration, followed by a phosphor screen (P43) and optical read-out using a lens (Kowa, LM35HC) and CMOS (FLIR, GS3-U3-23S6M-C) camera. A 2x2 median filter is applied to all electron beam images to avoid dead pixels. 

\begin{figure}[h]
   \centering      
   \includegraphics[width=8.0cm]{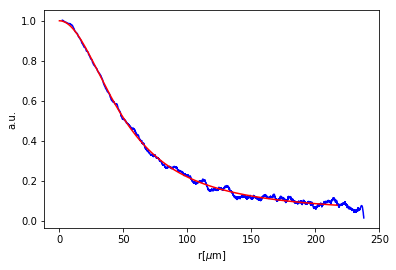}
  \caption{Normalized radial intensity distribution obtained from 31 single electron detection events. The data points (10974) were smoothed out using a Savitzky-Golay filter with a window size of 601 and a polynomial order of 3. The fitted Voigt profile represents a convolution of a Gaussian profile with a FWHM of $\SI{20.4}{\mu m}$, and a Lorentzian profile with a FWHM of $\SI{95.2}{\mu m}$.}
 \label{fig:PSF_fit}
\end{figure}

To characterize the point-spread function of our detector, we recorded 31 single electron detection events at the gain settings that were used in the experiments. We first fit the detection events with a Gaussian distribution to find their center. We then plot the radial distribution of counts (see Fig.~\ref{fig:PSF_fit}), which is well represented by a Voigt profile (red) with a FWHM of $(99.9\pm0.3)\mu$m .

\section{\label{sec:D}Ponderomotive Interaction}
\label{sec:Appendix_ponderomotive_interaction}
Here we derive the light-induced phase modulation of an electron beam in free space, using SI units throughout. 
We define the direction of propagation to be the z-axis and assume the electron waves travel at a relativistic mean velocity $\bm v = v\bm e_z$, which sets the Lorentz factor as $\gamma=1/\sqrt{1-v^2 / c^2}$ and the energy as $E_e = \gamma m_e c^2$. 
In such a setting, the electron wave function $\psi$ can be factorized into the rapidly oscillating plane-wave component at $E_e$ and a slowly varying envelope $\psi_{\perp}$ that accounts for the transverse wave profile,
\begin{equation}
    \psi(\bm{r},t) = \psi_{\perp}(\bm{r},t) e^{i\gamma m_e(v z - c^2 t)/\hbar}.
\label{eq:initialelectronwavefunction}
\end{equation}
Assuming that light-induced variations of the electron energy be small compared to the initial energy $E_e$ (called non-recoil approximation), it is shown in~\cite{PhysRevLett.126.123901} that the Dirac equation for the electron state reduces to the effective Schrödinger equation 
\begin{equation}\label{eq:effSE}
    (\partial_t +\bm{v}\cdot \nabla)\psi_{\perp}(\bm{r},t) = -\frac{i}{\hbar} U (\bm{r},t) \psi_{\perp}(\bm{r},t),
\end{equation}
with the effective interaction potential
\begin{eqnarray}
    U(\bm r,t) &=& \frac{e^2}{2m_e \gamma} \left[ A_x^2 (\bm r, t) + A_y^2 (\bm r, t) + \frac{1}{\gamma^2} A_z^2 (\bm r, t) \right] \nonumber \\
    && + e \bm{v} \cdot \bm{A}(\bm r, t) - e \Phi(\bm r, t). \label{eq:intPotential}
\end{eqnarray}
Here, $\bm{A}$ and $\Phi$ are the real-valued vector and the scalar potential of the light field, respectively. By integrating Eq.~\eqref{eq:effSE}, we find that the evolution of the state $\psi_{\perp}$ between two times $t_0$ and $t_1$ can be expressed as
\begin{eqnarray} \label{eq:scatteringTrafo}
    \psi_{\perp}(\bm{r}+\bm v t_1,t_1) &=& \psi_{\perp}(\bm{r}+\bm{v}t_0,t_0) \\
    && \times  \exp \left[-\frac{i}{\hbar} \int_{t_0}^{t_1} \diff t' \, U(\bm{r}+ \bm{v}t', t') \right]. \nonumber 
\end{eqnarray}
The free propagation amounts to a position shift while the light field contributes an additional phase modulation. In fact, if we choose a sufficiently large interval $(t_0,t_1) \to (-\infty,\infty)$ that covers the whole duration of the electron-light interaction, Eq.~\eqref{eq:scatteringTrafo} can be understood as the instantaneous scattering transformation from the incoming to the outgoing electron wave evaluated at the same time,
\begin{equation}\label{eq:Eikotrafo}
    \psi_{\perp}^{\text{(out)}} (\bm r,t) = e^{i\varphi (\bm r)}\psi_{\perp}^{\text{(in)}} (\bm r,t), 
\end{equation}
with the scattering phase $\varphi (\bm r) = - \int \diff t'\, U(\bm r + \bm v t',t')/\hbar$. The time coordinate $t$ marks the scattering event, i.e.~the point in time when the electron wave packet hits the light pulse.

Regarding the interaction potential from Eq.~\eqref{eq:intPotential}, notice that we may fix the gauge such that $\Phi=0$ and we can safely neglect the term $\bm v \cdot \bm A$ as we integrate the optical field over a pulse length of $10^2$ periods~\cite{PhysRevLett.126.123901}. Hence we may omit the respective terms in $U$. 

Given a paraxial light pulse (as focused by moderate numerical apertures) with central frequency $\omega_L$ that propagates in the $\pm z$-direction and is linearly polarized along $x$, the associated complex-valued electric field can be written as
\begin{equation}
    \bm E (\bm r, t) = \mathcal{E}_0 \bm e_x g(x,y) e^{-i\omega_L (t \mp z/c)} u \left(t \mp \frac{z}{c} \right),
\label{eq:efieldparaxial}
\end{equation}
with $\mathcal{E}_0$ the characteristic field strength, $g(x,y)$ the transverse mode profile, and $u(t)$ the temporal pulse envelope. For simplicity, these three terms are taken to be real-valued in the following. The real-valued vector potential in Eq.~\eqref{eq:intPotential} is related to the physical field amplitude via $-\partial_t \bm A (\bm r,t) = \Re \{ \bm E (\bm r,t) \}$.

For the temporal pulse width considered here ($\Delta t = \SI{280}{fs}$), the envelope $u(t)$ varies on much greater time scales than the optical oscillation period ($2\pi/\omega_L=\SI{3.5}{fs}$), which allows us to write the vector potential to leading order as  
\begin{eqnarray} \label{eq:physicalvectorpotential}
    \bm A (\bm r ,t) &=& \Re\left\{\frac{\bm E (\bm r,t)}{i\omega_L}\right\} \\
    &=& - \frac{\mathcal{E}_0 \bm e_x}{\omega_L} g(x,y) \sin\left[\omega_L\left(t \mp \frac{z}{c}\right)\right] u\left(t \mp \frac{z}{c}\right) \nonumber
\end{eqnarray}
Plugging this into Eq.~\eqref{eq:intPotential}, the scattering phase becomes
\begin{eqnarray}
    \varphi_v(\bm{r})  &=& -\frac{e^2 \mathcal{E}_0^2 g^2(x,y)}{2 m_e \gamma \hbar \omega_L^2} \int \diff t' \, u^2\left(t' \mp \frac{ z+vt'}{c}\right) \nonumber \\
    && \times \sin^2\left[\omega_L\left( t' \mp \frac{z+vt'}{c}\right)\right] \\
    &=& -\frac{e^2 \mathcal{E}_0^2 g^2(x,y)}{2 m_e \gamma \hbar \omega_L^2 (1\mp \beta)} \int \diff t \, u^2(t) \sin^2 (\omega_L t) . \nonumber 
\end{eqnarray}
In the last line, we substitute $t' \mp (z+vt')/c \to t$ and introduce $\beta = v/c$. As a next step, we can write $\sin^2(\omega_L t)=[1-\cos(2\omega_L t)]/2$ and neglect the integral over $u^2(t)\cos(2\omega_L t)$ due to the slowly varying pulse envelope, which leaves us with the ponderomotive scattering phase averaged over the fast optical oscillation,
\begin{equation}
    \varphi_v(x,y) \approx -\frac{e^2 \mathcal{E}_0^2 g^2(x,y)}{4 m_e \gamma \hbar \omega_L^2 (1\mp \beta)} \int \diff t \, u^2(t). \label{eq:pondPhase}
\end{equation}
Notice that the phase no longer depends on the $z$-coordinate, which is consistent with the paraxial regime for both light and electron waves.

Finally, we can express the laser field amplitude in terms of a more directly accessible parameter. To this end, consider the field energy flowing through the $xy$-plane per unit area and time, as described by the Poynting vector $\bm S = \Re \{ \bm E \} \times \Re \{ \bm H \}$ with $\bm H = c\varepsilon_0 \bm e_z \times \bm E$ to leading order. The measured laser pulse energy $E_L$ is obtained by taking the integral of $|\bm S|$ over the plane and over time, which leads to
\begin{eqnarray}
    E_L &=& c\varepsilon_0 \mathcal{E}_0^2 \int \diff t \, u^2(t) \cos^2 (\omega_L t)  \int \diff x\diff y \, g^2 (x,y) \nonumber \\
    &\approx& \frac{c\varepsilon_0}{2} \mathcal{E}_0^2 \int \diff t \, u^2(t) \int \diff x\diff y \, g^2 (x,y),
\end{eqnarray}  
assuming once again a slow temporal pulse profile. We can now eliminate this profile and the field strength in the ponderomotive phase (Eq.~\eqref{eq:pondPhase}) and simplify
\begin{equation}
    \varphi_v (x,y) =-\frac{\alpha}{2\pi(1 \mp \beta)} \frac{E_L}{E_e} \frac{\lambda_L^2 g^2(x,y)}{\int \diff x\diff y \, g^2(x,y)}.
\label{eq:phaseimprintlabframe}
\end{equation}
This equals the expression given in the main text, with $\alpha$ the fine structure constant and $\lambda_L = 2\pi c/\omega_L$ the pulse wavelength. The transverse light beam profile $g^2(x,y)$ is divided by its integral and can thus be given in arbitrary units. Moreover, the phase is Lorentz covariant since it reads the same in the electron rest frame where $\beta'=0$, $E_e'=m_e c^2$, $\lambda_L' = \lambda_L \sqrt{(1 \pm \beta)/(1 \mp \beta)}$, and $E_L' = E_L \sqrt{(1 \mp \beta)/(1 \pm \beta)}$.

\section{\label{sec:E}Free electron propagation}

The wavefunction of a freely propagating relativistic electron can be expanded in terms of plane-wave bispinor solutions to the Dirac equation \cite{messiah2014quantum}. Given the momentum $\bm p$ and the spin-$1/2$ state $|s\rangle$, the solution reads as
\begin{equation}
    \Psi_{\bm p,s} (\bm r,t) = \frac{1}{\sqrt{\mathcal{N}_p}} e^{i(\bm p \cdot \bm r - E_p t)/\hbar} \begin{pmatrix} |s\rangle \\ \frac{c}{E_p + m_e c^2} \bm p \cdot \hat{\bm{\sigma}}|s\rangle    \end{pmatrix},
\end{equation}
with $E_p = \sqrt{m_e^2 c^4 + p^2c^2}$ the energy, $\mathcal{N}_p = \sqrt{(E_p+m_ec^2)/2E_p}$ a normalization factor, and $\hat{\bm{\sigma}} = (\hat \sigma_x,\hat \sigma_y,\hat \sigma_z)$ the Pauli matrices. The electronic character of the solution ($E_p > 0$) and the spin state are preserved under the free evolution. Hence, a spatial wave packet of fixed spin can be expressed in terms of a scalar field $\psi(\bm r,t)$,
\begin{eqnarray}
    \Psi_s (\bm r,t) &=& \int\diff^3 p \, c(\bm p) \Psi_{\bm p,s} (\bm r,t) = \begin{pmatrix} \psi(\bm r,t) |s\rangle \\ -i \hat{\bm{\sigma}}\cdot \nabla \chi(\bm r,t) |s\rangle \end{pmatrix} \nonumber \\
    \psi(\bm r,t) &=& \int\diff^3 p\, \frac{c(\bm p)}{\sqrt{\mathcal{N}_p}} e^{i(\bm p \cdot \bm r - E_p t)/\hbar}, \nonumber \\
    \chi(\bm r,t) &=& \int\diff^3 p\, \frac{c(\bm p)}{\sqrt{\mathcal{N}_p}} \frac{\hbar c}{E_p + m_ec^2} e^{i(\bm p \cdot \bm r - E_p t)/\hbar}.
\end{eqnarray}
Note that the (unnormalized) positronic field component $\chi$ can be obtained from the scalar wavefunction $\psi$ via $(i\hbar \partial_t + m_e c^2)\chi = \hbar c \psi$. The electronic wave function $\psi$ obeys the Klein-Gordon equation, 
\begin{equation}
    \left( \hbar^2 \partial_t^2 - \hbar^2 c^2 \nabla^2 + m_e^2 c^4 \right) \psi = 0.
\end{equation}
Here, we consider paraxial electron waves with a relativistic forward velocity $v$ along the $z$-axis, such that $E_p = E_e = \gamma m_e c^2$ and $p = \gamma m_e v$. Replacing $\hbar^2\partial_t^2 \to -E_e^2$ in the Klein-Gordon equation, we arrive at the Helmholtz wave equation, $(\nabla^2 + k^2)\psi = 0$, with the effective wave number $k = \sqrt{E_e^2 - m_e^2c^4}/\hbar c = \gamma m_e v/\hbar$. Propagation of the electron waves through the apertures of our experiment can thus be described using Fresnel diffraction theory.

We remark that, for the numerical simulation of electron pulses in our experiment, we ignore the spread of forward velocities around the mean value $v$. Its impact on the results is negligible compared to other uncertainties such as the finite resolution in the detector.

\subsection{Wave optics simulations}

In order to verify the measured light lensing effect, we compute the Fresnel diffraction image of a paraxial electron wave of forward velocity $v$, after it interacted with a light pulse via the ponderomotive potential. Omitting constant prefactors and assuming a circular aperture of radius $R$ in the interaction plane, the electron density distribution in the detection plane at the effective distance $d$ reads as
\begin{equation}
    I(\bm r_S) \propto \left| \int_{|\bm r| \leq R} \!\!\!\diff^2 r \, \psi_{\perp}^{\text{(in)}} (\bm r) e^{i\varphi(\bm r) + ik (\bm r^2 - 2\bm r \cdot \bm{r}_S)/2d} \right|^2.
\end{equation}
For a circularly symmetric ponderomotive phase $\varphi (\bm r) = \varphi(\rho)$ and for a centered and converging incident wave $\psi_{\perp}^{\text{(in)}} \propto \exp(-ik\rho^2/2d_c)$, with $\rho = \sqrt{x^2+y^2}$, we can switch to polar coordinates and simplify 
\begin{eqnarray}
    I(\rho_S) &\propto& \left| \int_{-\pi}^\pi \!\!\! \diff \theta \int_0^R \!\!\! \rho \diff \rho \, e^{ik\rho [\rho(1-d/d_c) - 2\rho_S \sin\theta ]/2d + i\varphi(\rho)} \right|^2 \nonumber \\
    &\propto& \left| \int_{-\pi}^\pi \!\!\! \diff \theta \int_0^1 \!\!\! \diff \xi \, e^{i\kappa\delta \xi -i\kappa \sqrt{\xi} \sin\theta \rho_S/R + i\Phi(\xi)}\right|^2. \label{eq:FresnelSignal}
\end{eqnarray}
Here, we have substituted $\xi = (\rho/R)^2$ and introduced $\Phi(\xi) = \varphi(R\sqrt{\xi})$ and the dimensionless parameters $\kappa = kR^2/d$ and $\delta = (d_c-d)/2d_c$. Carrying out the angular integral would yield the Bessel function $J_0 (\kappa \sqrt{\xi} \rho_S/R)$, and the remaining $\xi$-integral could be computed numerically. However, we have $\kappa,|\delta| \gg 1$ and a strongly varying phase $\Phi$ in our setting, which renders direct numerical integration methods unstable. 

A viable alternative is to first perform a stationary phase approximation of the $\xi$-integral in Eq.~\eqref{eq:FresnelSignal} and then compute the remaining $\theta$-integral numerically. To this end, let $F(\xi)$ denote the total complex phase under the integral and $\xi_n \in (0,1)$ be all non-degenerate stationary points such that $F'(\xi_n)=0$ and $F''(\xi_n) \neq 0$. Explicitly, the points depend on $\theta$ and are solutions to 
$ 2\sqrt{\xi_n} [\kappa\delta + \Phi'(\xi_n)] = \kappa \sin\theta \rho_S / R$.
We may then approximate
\begin{equation}\label{eq:statPhase}
    \int_0^1 \!\!\! \diff\xi \, e^{iF(\xi)} \approx \sum_n \sqrt{\frac{2\pi i}{F''(\xi_n)}} e^{i F(\xi_n)}.
\end{equation}
For the wave optics simulations of the Gauss and the Laguerre-Gauss light pulses in the main text, we had a numerical solver find $\xi_n \in (0,1)$ for each relevant value of $\sin\theta \rho_S/R$, compute Eq.~\eqref{eq:statPhase}, and finally evaluate the $\theta$-integral. The resulting radial diffraction pattern $I(\rho_S)$ was then mapped onto a two-dimensional grid of $2001 \times 2001$ detector pixels and smeared over the finite detector pixel size as described in Sec. C.

\subsection{Raytracing simulations}

For direct comparison, we compute the light-induced electron lensing using the approximation of geometric optics. This omits diffraction effects and treats the electrons as ballistic particles moving on piecewise rectilinear trajectories (rays).

Each ray is initialized in the interaction plane with its transverse position and its angle with respect to the optical axis: $\eta_0^\top=(x_0 ~ y_0 ~ \theta_{x,0} ~ \theta_{y,0})$. For a convergent probe, the angles are given by $\theta_{x, 0} = x/d_c$ and $\theta_{y, 0} = y/d_c$ where $d_c$ is the distance from the interaction plane to the electron crossover.
The interaction with the ponderomotive potential changes the angle in $x$-direction as $\theta_{x,1}=\theta_{x,0}+\hbar \nabla_{x} \varphi(x,y)/p_z$, and analogously in $y$-direction.
Here, $p_z$ is the longitudinal momentum.
Interaction induced changes in $p_z$ can be neglected.
The positions $x_0$ and $y_0$ remain unchanged in the thin optics approximation.
Free space propagation to the screen (distance $d$) is calculated as $\eta_{s}=S\eta_1$, where $S$ is the free-space propagation matrix: 
\begin{equation}
S =
    \begin{pmatrix}
        1 & 0 & d & 0 \\
        0 & 1 & 0 & d \\
        0 & 0 & 1 & 0 \\
        0 & 0 & 0 & 1 \\
    \end{pmatrix}
\end{equation}
The raytracing simulation in Fig.~\ref{fig:temporal}(c)(top) was done using the Monte-Carlo method with $5\times10^5$ randomly sampled starting conditions. 

For the perfect lens calculations shown in Fig.~\ref{fig:lensing}(b), the ponderomotive interaction is described analytically using the ABCD matrix of a thin lens $L$:
\begin{equation}
L =
    \begin{pmatrix}
        1 & 0 & 0 & 0 \\
        0 & 1 & 0 & 0 \\
        -1/f & 0 & 1 & 0 \\
        0 & -1/f & 0 & 1 \\
    \end{pmatrix},
\end{equation}
where the focus length can be calculated according to Eq.~(\ref{eqn:focal_length_laguerre}) and Eq.~(\ref{eqn:focal_length_gaussian}). The electron trajectories at the screen are then described as $\eta_{s}=SL\eta_0$. We get an analytic expression for the radius of the beam on the screen, 
\begin{align}
        \rho_{s}(E_L) &= \rho_0 \left( 1-\frac{d}{f(E_L)} \right) + d\theta_{\rho,0} ,
    \label{eqn:spot_size}
\end{align}
where $\rho_0$ and $\theta_{\rho,0}$ are defined by the outermost incoming beam. 

Our simulation results in the main text agree well with both the measured images and the Fresnel wave diffraction results.

\section{\label{sec:F}Lens calculation}

We now consider the ponderomotive phase of circularly symmetric light pulses in order to realize electron lenses.
An ideal lens for electrons of wavelength $\lambda_e$ would impart a phase modulation to the electron wave that is quadratic in the distance $\rho = \sqrt{x^2+y^2}$ to the optical axis, 
\begin{align}
    \varphi_{\text{lens}}(\rho) = - \frac{\pi}{f \lambda_e} \rho^2 .
    \label{eqn:lens_phase_shift}
\end{align}
Whereas the focal length $f$ is positive for convex lenses and negative for concave lenses.

\subsection{Convex lens}

The negative quadratic phase modulation of a convex lens can be realized in the center of a focused Laguerre-Gauss mode with radial index $p=0$ and azimuthal index $l=1$ (LG10). The spatial intensity profile of LG10 is then given by
\begin{align}
    g^2_{10}(\rho) =  \frac{\rho^2}{w_{10}^2}  \exp \left( -2\rho^2/w_{10}^2 \right),
    \label{eqn:LG10}
\end{align}
where $w_{10}$ is the beamwaist of the Laguerre-Gauss mode. Using a Taylor expansion around $\rho/w_{10}=0$, we get the desired $\rho^2$-dependence with $g^2_{10} \approx \rho^2/w_{10}^2 - \mathcal{O}(\rho^4/w_{10}^4)$.

\begin{figure}[t]
    \centering
    \includegraphics[width=8.5cm, height=8.5cm]{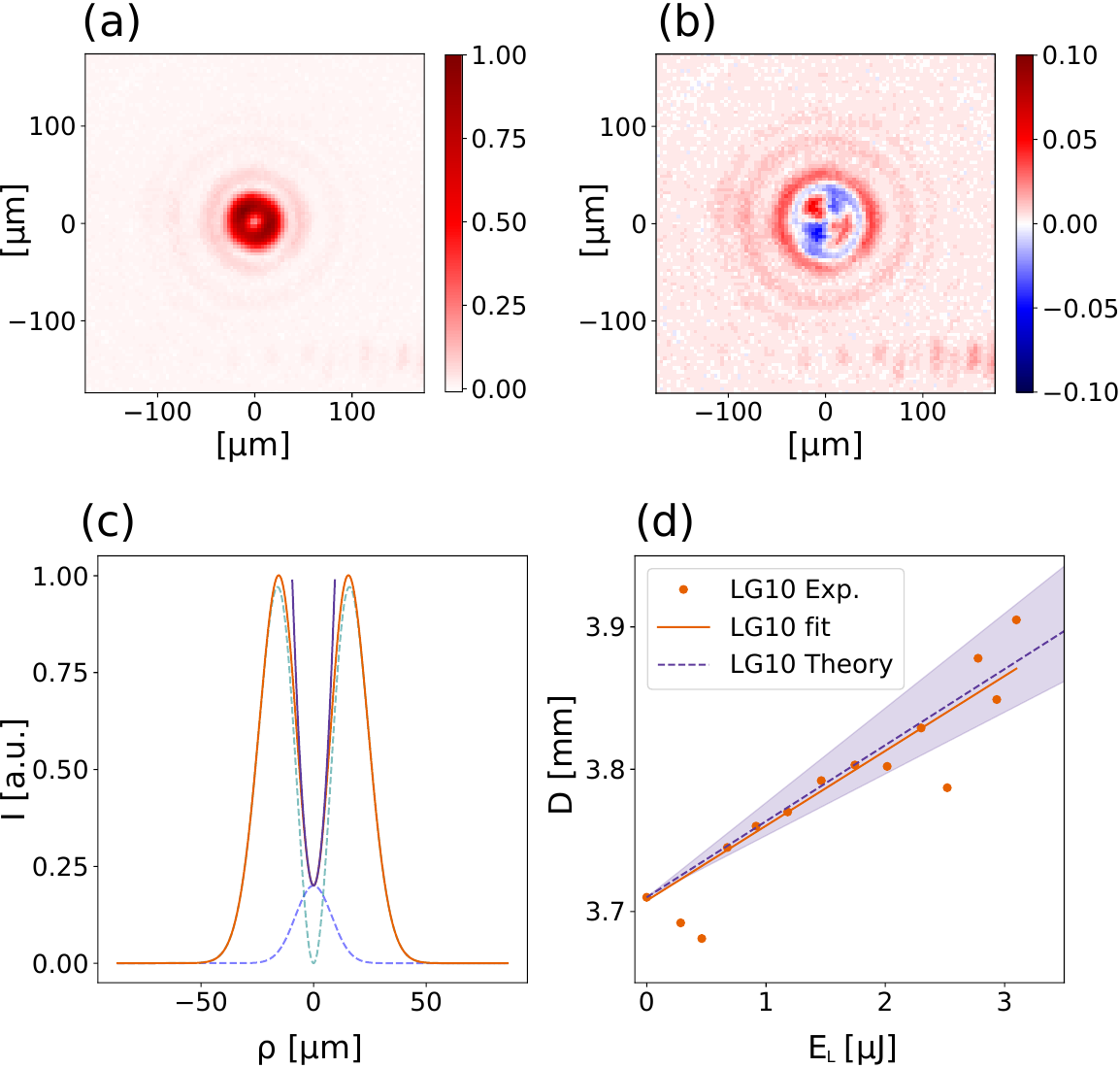}
    \caption{(a) Spatial intensity distribution of the LG10 beam measured in the conjugate plane. (b) Residuals of a fit to the distribution. The fit function is a sum of a dominant LG10 mode and a minor LG00 mode. (c) Crosscut showing the fit (orange), its LG10 (turquoise) and LG00 (blue) component, as well as a the quadratic Taylor expansion around $\rho=0$ (violet). (d) Diameter of electron beam at the detector as a function of laser pulse energy, along with the theory and an error region due to our error in determining the waist of the dominant LG10 component.}
    \label{fig:LG10_int_cross_section}
\end{figure}

We can now calculate the focal length of the ponderomotive electron lens. We insert Eq.~(\ref{eqn:LG10}) into the denominator in Eq.~(\ref{eq:phaseimprintlabframe}) and its Taylor expansion into the numerator, and compare the coefficient of the quadratic term with Eq.~(\ref{eqn:lens_phase_shift}). Eventually we get
\begin{align}
    f_{10}(E_L) &= (1\mp \beta) \frac{\pi^3 w_{10}^4 E_e}{2 \alpha \lambda_L^2 \lambda_e} \frac{1}{E_L}
    \label{eqn:focal_length_laguerre}
\end{align}

The experimentally recorded intensity distribution shown in Fig.~\ref{fig:LG10_int_cross_section}(a) is not a perfect LG10 mode. Most notably, the intensity does not go to 0 in the center, and higher diffraction orders from the SLM make up for a significant fraction of the measured pulse energy. This is taken into account by fitting the distribution with the sum of a dominant LG10 and a minor LG00 mode for the intensity observed in the center. The residuals of this fit are shown in Fig.~\ref{fig:LG10_int_cross_section}(b) and amount to up to $5.3\%$ of the peak intensity, and less than $2.6\%$ at the central part of the beam which is illuminated by the electrons. The fitted distribution makes up for $56\%$ of the total laser power. Figure~\ref{fig:LG10_int_cross_section}(c) shows a crosscut through the fit and its LG10 and LG00 components, along with a quadratic fit to the center of the distribution. The quadratic coefficient is dominated by the LG10 component, for which we get a waist of $w_{10}=\SI{4.51(22)}{\mu m}$. In the experiment, the maximum energy within the fitted distribution is $E_L=\SI{3.1}{\mu J}$, for which we get a focal length of $f=\SI{4.5}{mm}$. Figure~\ref{fig:LG10_int_cross_section}(d) shows the measured diameter of the electron beam on the detector. Due to the asymmetry in the light distribution that is visible in the residuals, slightly cylindrical lensing is observed in the experiment. Both here, and in the calculation of the focal length, we use the geometric mean between the fitted minor and major axis of the electron beam distribution at the MCP, which we deduced using the Otsu segmentation method in the image processing tool Fiji~\cite{Schindelin2012}.

Figure~\ref{fig:LG10_int_cross_section}(d) also shows a linear fit to the data (orange), as well as the theoretical expectation from Eq.~\eqref{eqn:focal_length_laguerre} (dashed line). 
Note also that our experimental $5\%$ error in determining the waist of the beam enters the equation to the fourth power, which yields the shaded area in the plot.
The two outliers at small pulse energies were not taken into account in the fit. These are most likely due to misalignment which proofed difficult at low pulse energies. Apart from these outliers, we see that the measurement became more unstable at higher pulse energies, most likely due to thermally induced misalignment. 
  
\subsection{Concave lens}

The positive quadratic phase distribution for a concave lens can be achieved with a Gaussian intensity distribution. The spatial intensity profile of the Gaussian beam is given by
\begin{align}
    g^2_{00}(\rho) =  \exp \left( -2\frac{\rho^2}{w_{00}^2} \right) \text{,}
    \label{eqn:gaussian_intensity_profile}
\end{align}
and its Taylor expansion around $\rho/w_{00}=0$ gives us $g^2_{00}(\rho) \approx 1-2\rho^2/w_{00}^2+\mathcal{O}(\rho^4/w_{00}^4)$.

\begin{figure}[h]
    \centering
    \includegraphics[width=8.5cm, height=8.5cm]{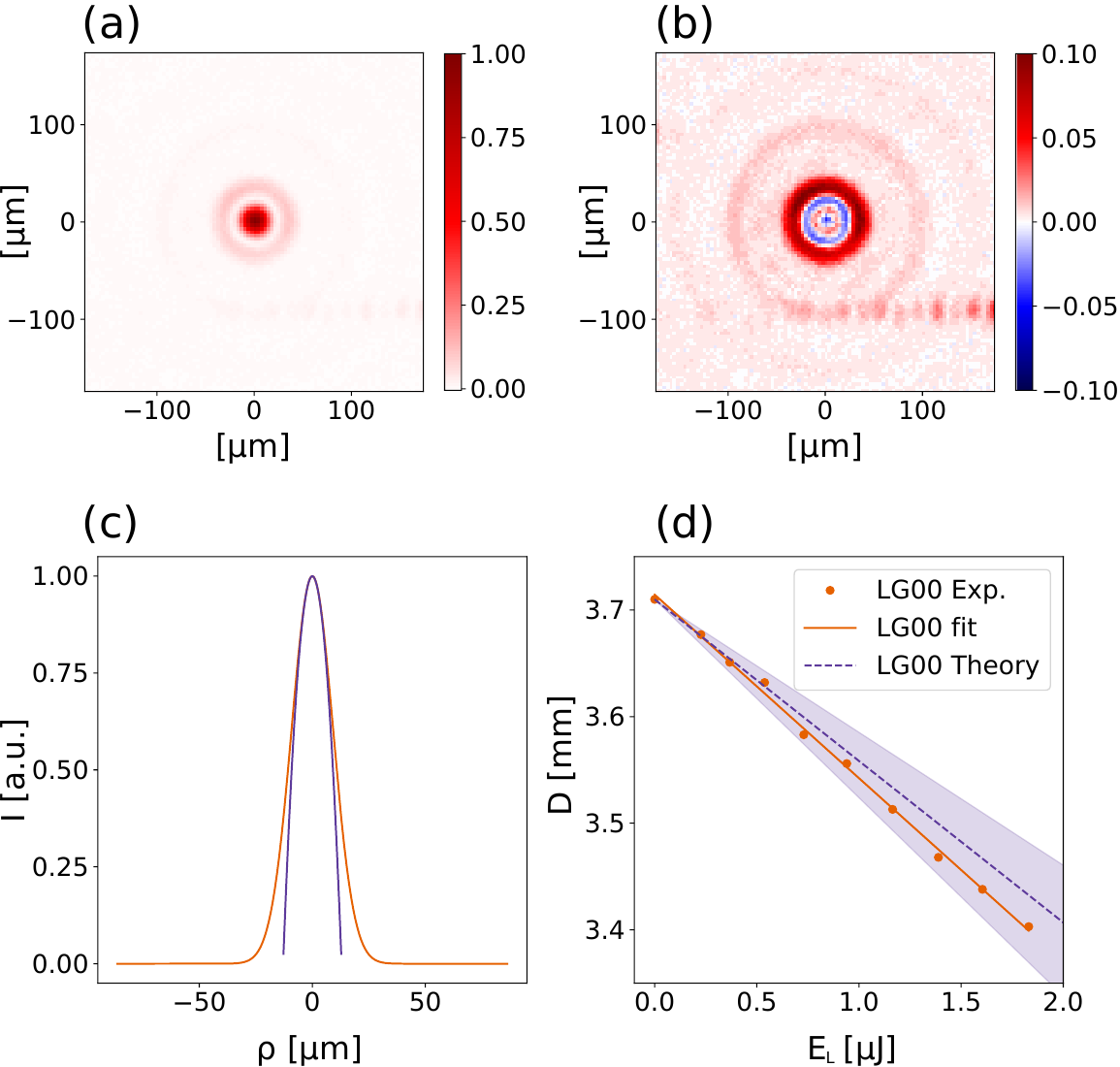}
    \caption{(a) Spatial intensity distribution of the LG00 beam measured in the conjugate plane. (b) Residuals of a LG00 fit to the distribution. (c) Crosscut showing the fit (orange), and its quadratic Taylor expansion around $\rho=0$ (violet). (d) Diameter of electron beam at the detector as a function of laser pulse energy, along with the theory and an error region due to our error in determining the waist of the LG00 mode.}
    \label{fig:gaussian_beam_plot}
\end{figure}

We calculate the phase modulation according to Eq.~(\ref{eq:phaseimprintlabframe}) and compare the coefficient of the quadratic term to retrieve the focal length of the electron lens
\begin{align}
    f(E_L) = -(1 \mp \beta) \frac{\pi^3 w_{00}^4 E_e}{2\alpha \lambda_L^2\lambda_e} \frac{1}{E_L}
    \label{eqn:focal_length_gaussian}
\end{align}

The experimentally recorded intensity distribution shown in Fig.~\ref{fig:gaussian_beam_plot}(a) is not a perfect Gaussian. 
Most notably, there's an outer diffraction ring, as well as higher diffraction orders from the SLM, that account for a significant amount of the measured laser pulse energy. 
This can be seen in the residuals of the Gaussian fit (Fig.~\ref{fig:gaussian_beam_plot}(b)). The electrons only hit the central part of the distribution, where the error is $<3\%$. 
The fitted Gaussian makes up for $45\%$ of the total laser power and has a width of $w_{00}=\SI{3.65(18)}{µm}$. Figure~\ref{fig:gaussian_beam_plot}(c) shows a crosscut of the Gaussian, along with a quadratic fit to it. In the experiment, the maximum energy within the Gaussian is $E_L = \SI{2.5}{µJ}$, resulting in a focal length of $f=\SI{-1.6}{mm}$. 
Figure~\ref{fig:gaussian_beam_plot}(d) shows the measured diameter of the electron beam on the detector, a linear fit to the data (orange), as well as the theoretical expectation according to Eq.~\eqref{eqn:focal_length_gaussian} (dashed line). 
Note that our experimental $5\%$ error in determining the waist of the beam enters the equation to the fourth power, which yields the shaded area in the plot.

\newpage
\bibliography{literatur}

\end{document}